\documentclass{PoS}

\usepackage{bm}
\usepackage[utf8]{inputenc}
\title{The Matrix Element Method as a tool for precision and accuracy}

\ShortTitle{The MEM as a tool for precision and accuracy}

\author{\speaker{Till Martini}, Manfred Kraus, Sascha Peitzsch, and Peter Uwer\\
        Humboldt-Universit{\"a}t zu Berlin, Institut f{\"u}r Physik, Newtonstra{\ss}e 15, 12489 Berlin, Germany\\
        E-mail: \email{Till.Martini@physik.hu-berlin.de}, \email{Manfred.Kraus@physik.hu-berlin.de}, \email{Sascha.Peitzsch@physik.hu-berlin.de}, \email{Peter.Uwer@physik.hu-berlin.de}}

\abstract{
The Matrix Element Method is a promising multi-variate analysis tool which offers an optimal approach to compare theory and experiment according to the Neyman-Pearson lemma. However, until recently its usage has been limited by the fact that only leading-order predictions could be employed. The imperfect approximation of the underlying probability distribution can introduce a significant bias into the analysis which requires a major calibration for the method when applied to parameter determination. Moreover, estimating theoretical uncertainties by scale variation may yield unreliable results. We present the extension of the Matrix Element Method to next-to-leading order in QCD applicable to LHC data defined by common jet algorithms. The accuracy gain is illustrated by simulating a top-quark mass determination from single top-quark events generated with \textsc{POWHEG}+\textsc{Pythia}. Additionally, the method's potential for BSM parameter determination is demonstrated by simulating the extraction of a CP-violating Top-Yukawa coupling from events of single top-quarks in association with a Higgs boson.
}

\FullConference{
European Physical Society Conference on High Energy Physics - EPS-HEP2019 -\\
			10-17 July, 2019\\
			Ghent, Belgium}

\begin{document}
\section{Introduction}
In recent years the level of precision achieved in experimental analyses as well as the accuracy of theoretical predictions has been steadily improving. Thus, allowing for more and more precise tests of the Standard Model (SM). However, no obvious signs for physics beyond the Standard Model (BSM) have been reported so far. In order to provide for thorough scrutiny, the most efficient analysis methods, e.g. multi-variate methods allowing to use most of the information contained in the data, have to be combined with state-of-the-art theoretical input. The so-called Matrix Element Method (MEM) constitutes such an efficient method by offering an optimal approach to compare data to theory based on the principle of Maximum Likelihood. However, the accuracy of theoretical predictions usable within the MEM has been limited so far to the leading order (LO) only. In order to benefit from the advantages of the MEM in search of even the smallest hints for New Physics at the LHC, it is mandatory that higher-order-QCD corrections are included in the MEM. In this article, we present an extension of the MEM to next-to-leading-order (NLO) accuracy and illustrate the performance of the MEM@NLO for two example applications. The article is organised as follows. In section~\ref{sect:MEM} the MEM is briefly reviewed and issues concerning its leading-order-only formulation are discussed. An extension of the accuracy of the likelihood calculation to NLO in QCD is presented in section~\ref{sect:diffjetxs}. The MEM@NLO is consecutively utilised to study the impact of parton shower effects on the analysis in section~\ref{sect:psmem}. In section~\ref{sect:bsmmem} the sensitivity of the MEM@NLO in a concrete example of BSM parameter extraction is studied. Finally, the conclusions are given in section~\ref{sect:concl}.

\section{The Matrix Element Method}\label{sect:MEM}
The Matrix Element Method is a multi-variate analysis method which can be used for signal searches as well as parameter estimation \cite{Kondo:1988yd,Kondo:1991dw}. Being based on the principle of Maximum Likelihood it allows for an unambiguous statistical interpretation of the inferred results \cite{cowan1998statistical}. At the same time, the MEM offers a transparent connection between theory and experiment since the likelihood function is calculated within Quantum Field Theory which relies on first principles. The probability density for observing a detector signature $\bm{x}$ (a collection of $k$ variables characterising a measured event with $n$ final-state objects, $k\leq 3n-4$ for a lepton collider and $k\leq 3n-2$ for hadronic collisions) is given in terms of the differential cross section $d\sigma$ 
\begin{equation}\label{eq:prob}
P(\bm{x})={1\over \sigma}\int d^k y\;{d^k\sigma\over dy_1\ldots y_k}\;W(\bm{x},\bm{y}).
\end{equation}
The transfer function $W(\bm{x},\bm{y})$ accounts for the unfolding of the observed signature in the detector $\bm{x}$ to the theoretically modelled final-state configuration in a certain point of the phase space $\bm{y}$. In order to ensure
$
\int d^k x\; P(\bm{x}) =1
$
the transfer function needs to be normalised to one.
Assuming a perfect detector amounts to setting 
$
W(\bm{x},\bm{y})=\delta(\bm{x}-\bm{y})
$
which is done in the following. In general, the theoretical prediction for the cross section is a function of certain model parameters $\bm{\Omega}$ rendering the probability density in Eq.~\ref{eq:prob} a function of the model parameters as well
$
\sigma \equiv \sigma(\bm{\Omega})\quad \Rightarrow \quad P(\bm{x})\equiv P(\bm{x}|\bm{\Omega}).
$
This allows for parameter determination via the method of Maximum Likelihood: For a given set of $N$ measured events $\{\bm{x}_i\},\; i=1,\ldots,N$, the joint probability for having observed this set is a function of $\bm{\Omega}$, hence a likelihood function
$
\mathcal{L}(\bm{\Omega}|\{\bm{x}_i\})=\prod\limits_{i=1}^{N}P(\bm{x}_i|\bm{\Omega}).
$
Maximising this likelihood function with respect to $\bm{\Omega}$ yields an estimator $\hat{\bm{\Omega}}$ for the model parameters. 

The use of the MEM in the context of high energy physics has been pioneered in top-quark measurements at the Tevatron (see, e.g., \cite{Abbott:1998dn,Abazov:2004cs,Abulencia:2006mi,Abazov:2009ii,Aaltonen:2009jj}) and since then it has been utilised as a powerful standard tool also in LHC analyses (see, e.g., \cite{Chatrchyan:2012xdj,Englert:2015dlp,Gritsan:2016hjl,Aad:2015upn,Khachatryan:2015ila,Aad:2015gra,Khachatryan:2015tzo}, because it facilitates optimal use of the available information by offering the most efficient inference. However, until recently the establishment of the MEM as {\it the} method to analyse data in the future has been hindered by the fact that the calculation of the likelihood function in perturbative QCD had been restricted to the Born approximation only. Since next-to-leading order QCD corrections are typically not small, their inclusion is mandatory to achieve a better approximation of the full theory with reduced theoretical uncertainties. Taking into account the emission of extra radiation in the real corrections allows for the first time to model jet structure and additional jet activity in coloured final states thereby enabling the description of kinematics which do not fit the leading-order picture. Furthermore, calculating predictions beyond the Born approximation is necessary in order to uniquely fix the renormalisation scheme which unambiguously defines the model parameters of the QFT. A theoretical accuracy at NLO and beyond is nowadays considered the standard for up-to-date calculations. Therefore, quite some effort has been put into incorporating NLO corrections in the MEM over the course of the last decade  \cite{Alwall:2010cq,Soper:2011cr,Soper:2012pb,Soper:2014rya,Prestel:2019neg,Campbell:2012cz,Campbell:2013hz,Baumeister:2016maz}. Recently, a general extension of the MEM to full NLO accuracy has been achieved \cite{Martini:2015fsa,Martini:2017ydu,Martini:2018imv,Kraus:2019qoq}.

\section{Differential jet cross sections at NLO accuracy}\label{sect:diffjetxs}
\begin{figure}[h!]
\begin{center}
  \includegraphics[width=0.25\textwidth]{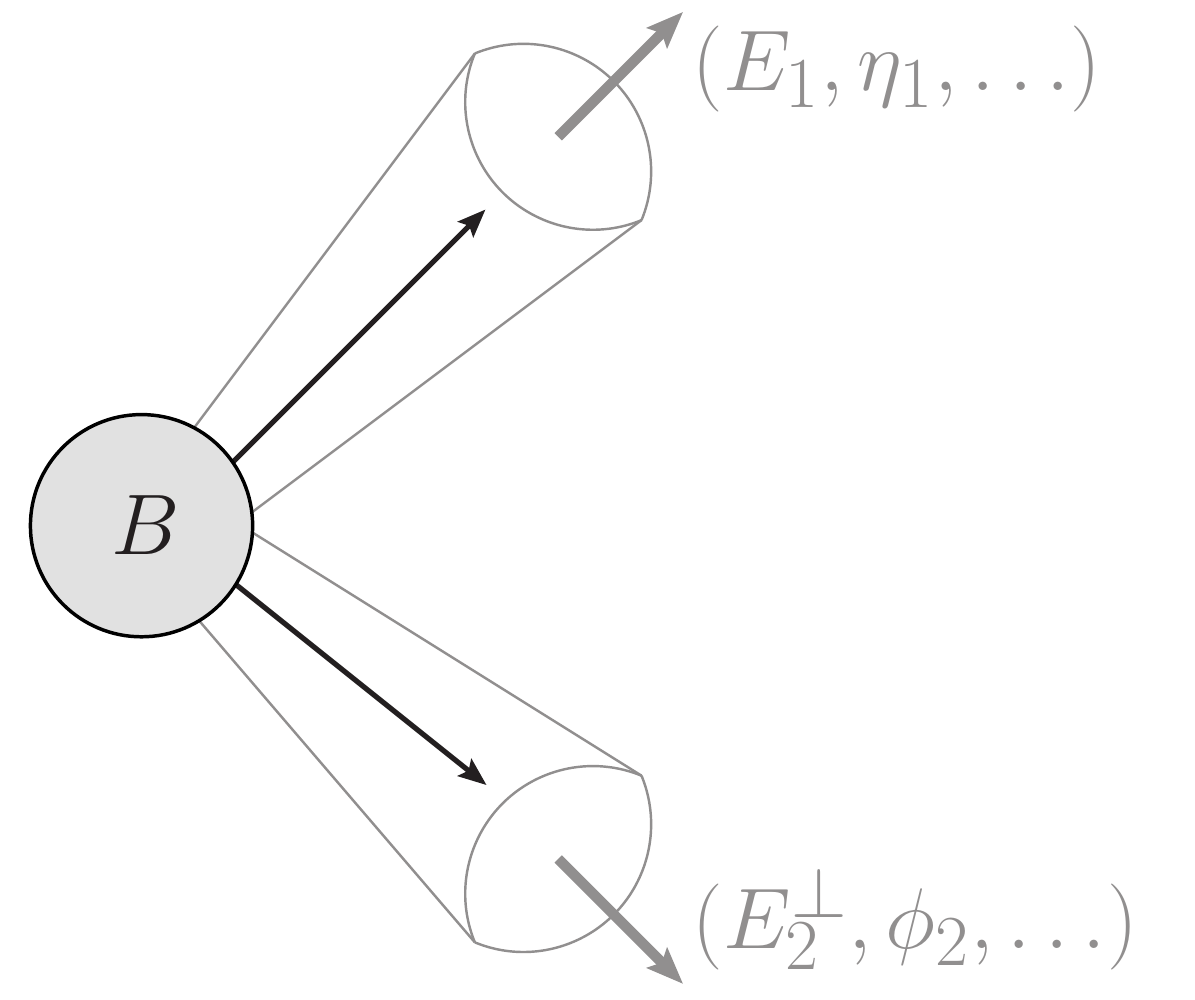}
  \includegraphics[width=0.25\textwidth]{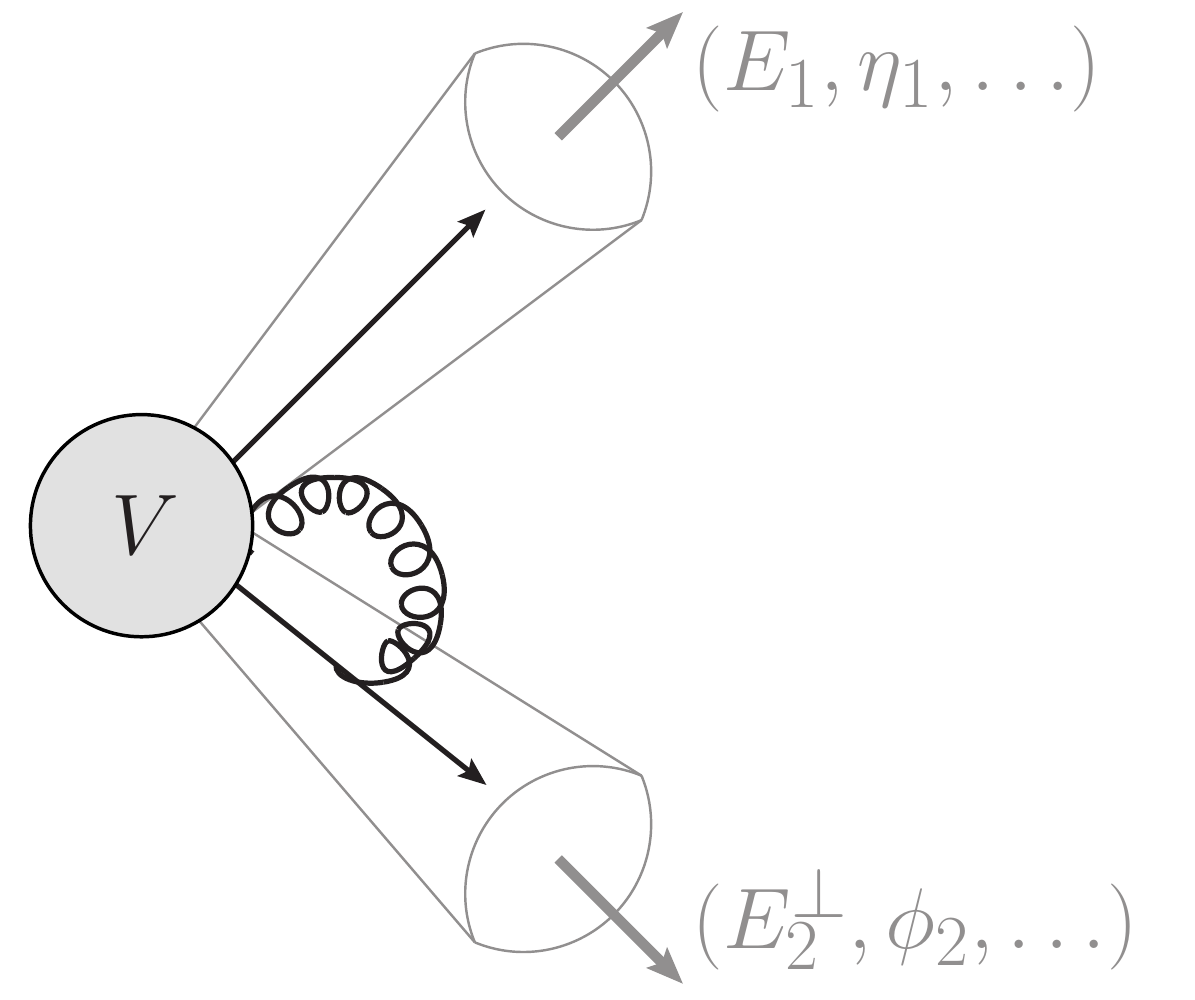}
  \includegraphics[width=0.25\textwidth]{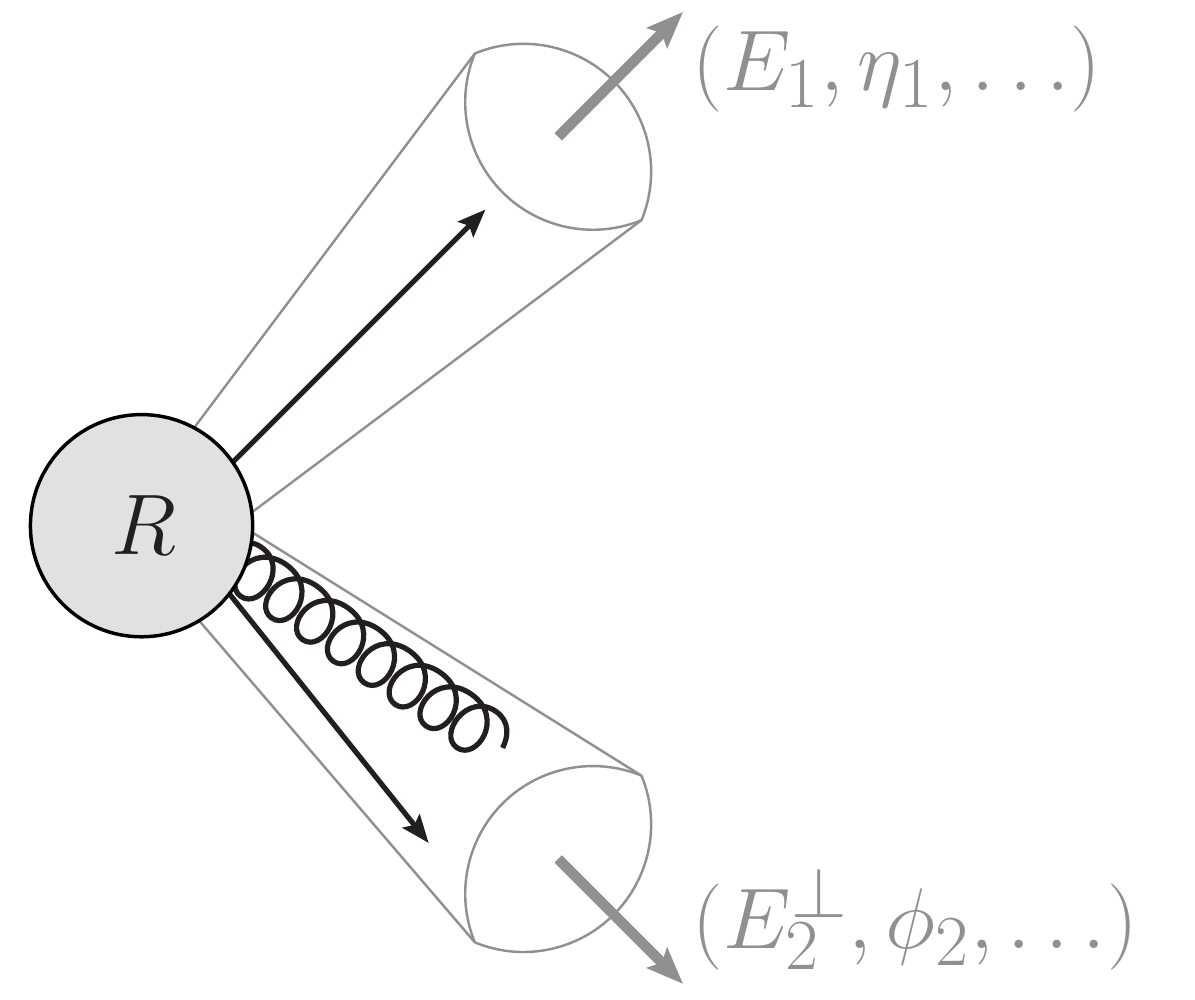}
\end{center}
\caption{Schematic representation of the Born approximation, virtual and real corrections contributing to a given final-state configuration defined by the jet variables $\bm{x}=(E_1,\eta_1,E^{\perp}_2,\phi_2,\ldots)$.
}
\label{fig:bvrjets}
\end{figure}

At the heart of the MEM lies the assignment of an event weight to the experimentally observed events according to the probability density given in Eq.~\ref{eq:prob}
\begin{equation}\label{eq:weight}
w(\bm{x})={1\over \sigma}{d^k\sigma\over dy_1\ldots y_k}.
\end{equation}
To make contact between theory and experiment we use jets as a well defined interface. The jets are obtained by applying a jet algorithm and may involve the recombination of primary objects (tracks of hadrons in the experimental setup, parton momenta in the theoretical description). Thus, the starting point is a set of fixed values of $k$ jet variables, e.g, $\bm{x}=(E_1,\eta_1,E^{\perp}_2,\phi_2,\ldots)$, for which the differential jet cross section has to be defined at NLO accuracy
\begin{equation}\label{eq:diffxs}
{d^k\sigma^{\mathrm{NLO}}\over dE_1\;d\eta_1\;dE^{\perp}_2\;d\phi_2 \ldots}={d^k\sigma^{\mathrm{B}}\over dE_1\;d\eta_1\;dE^{\perp}_2\;d\phi_2 \ldots}+{d^k\sigma^{\mathrm{V}}\over dE_1\;d\eta_1\;dE^{\perp}_2\;d\phi_2 \ldots}+{d^k\sigma^{\mathrm{R}}\over dE_1\;d\eta_1\;dE^{\perp}_2\;d\phi_2 \ldots}.
\end{equation}
Here, $\sigma^\mathrm{B}$ stands for the Born contribution while $\sigma^\mathrm{V}$ and $\sigma^\mathrm{R}$ denote the virtual and real corrections respectively.
The Born contribution is IR finite, e.g. free of soft and collinear divergences, and the mapping from partonic to jet momenta is given by trivial identification. However, the virtual and real corrections are separately IR divergent and when extra radiation is clustered by the jet algorithm the jet momenta become non-trivial functions of the partonic momenta (cf. Fig~\ref{fig:bvrjets}).
For IR-safe observables like jet variables the Kinoshita-Lee-Nauenberg theorem \cite{Kinoshita:1962ur,Lee:1964is} ensures that the separate IR-divergences cancel in the sum of virtual and real contributions. Thus, a $1$-to-$1$ correspondence of the virtual and real corrections is required enabling the efficient integration of unresolved partonic configurations in the real corrections at the same time. This can be achieved if the Lorentz-invariant $n$-particle phase space for the Born and virtual contributions can be parameterised in terms of the jet variables
\begin{displaymath}
d\mathrm{LIPS}_n=dE_1\;d\eta_1\;dE^{\perp}_2\;d\phi_2\ldots\times{d\mathrm{LIPS}_n\over dE_1\;d\eta_1\;dE^{\perp}_2\;d\phi_2 \ldots}.
\end{displaymath}
At the same time the $n+1$-particle phase space for the real corrections has to be factorised into the $n$-particle phase space times a phase space measure corresponding to the unresolved configurations
\begin{displaymath}
d\mathrm{LIPS}_{n+1}=dE_1\;d\eta_1\;dE^{\perp}_2\;d\phi_2\ldots\times{d\mathrm{LIPS}_n\over dE_1\;d\eta_1\;dE^{\perp}_2\;d\phi_2 \ldots}\times d\Phi_{\mathrm{unres}}.
\end{displaymath}
In \cite{Martini:2018imv,Kraus:2019qoq} is has been shown that this is possible for sensible choices of the jet variables: In particular, the jet variables cannot allow to reconstruct the invariant jet masses or the overall transverse momentum of the final state since this would prevent the mapping from real to Born-like kinematics which is required for the mutual cancellation of the IR singularities. With the phase space parameterisations sketched above and presented in detail in \cite{Martini:2018imv,Kraus:2019qoq} virtual and real corrections in Eq.~\ref{eq:diffxs} can be unambiguously combined and the unresolved configurations can be integrated out efficiently. This allows to define a weight at NLO accuracy for events given in terms of the jet variables
\begin{displaymath}
w(E_1,\eta_1,E^{\perp}_2,\phi_2,\ldots)={1\over \sigma^{\mathrm{NLO}}}\;{d^k\sigma^{\mathrm{NLO}}\over dE_1\;d\eta_1\;dE^{\perp}_2\;d\phi_2 \ldots}.
\end{displaymath}
This weight can be used to generate events $(E_1,\eta_1,E^{\perp}_2,\phi_2,\ldots)$ distributed according to the NLO prediction or in the MEM---extending it to NLO accuracy. In the next two sections two sample applications of the MEM@NLO are presented.

\section{Studying parton-shower effects in the MEM@NLO}\label{sect:psmem}
As a first example application of the MEM at NLO accuracy the impact of missing higher-order effects in the likelihood calculation is studied. To that end, single top-quark events are generated with \textsc{POWHEG} \cite{Alioli:2010xd,Alioli:2009je} and subsequently showered with \textsc{Pythia8} \cite{Sjostrand:2014zea}. Analysing these events with the MEM based on fixed-order calculations allows to single out how the parton shower affects the analysis. 
\begin{figure}[h!]
\begin{center}
  \includegraphics[width=0.497\textwidth]{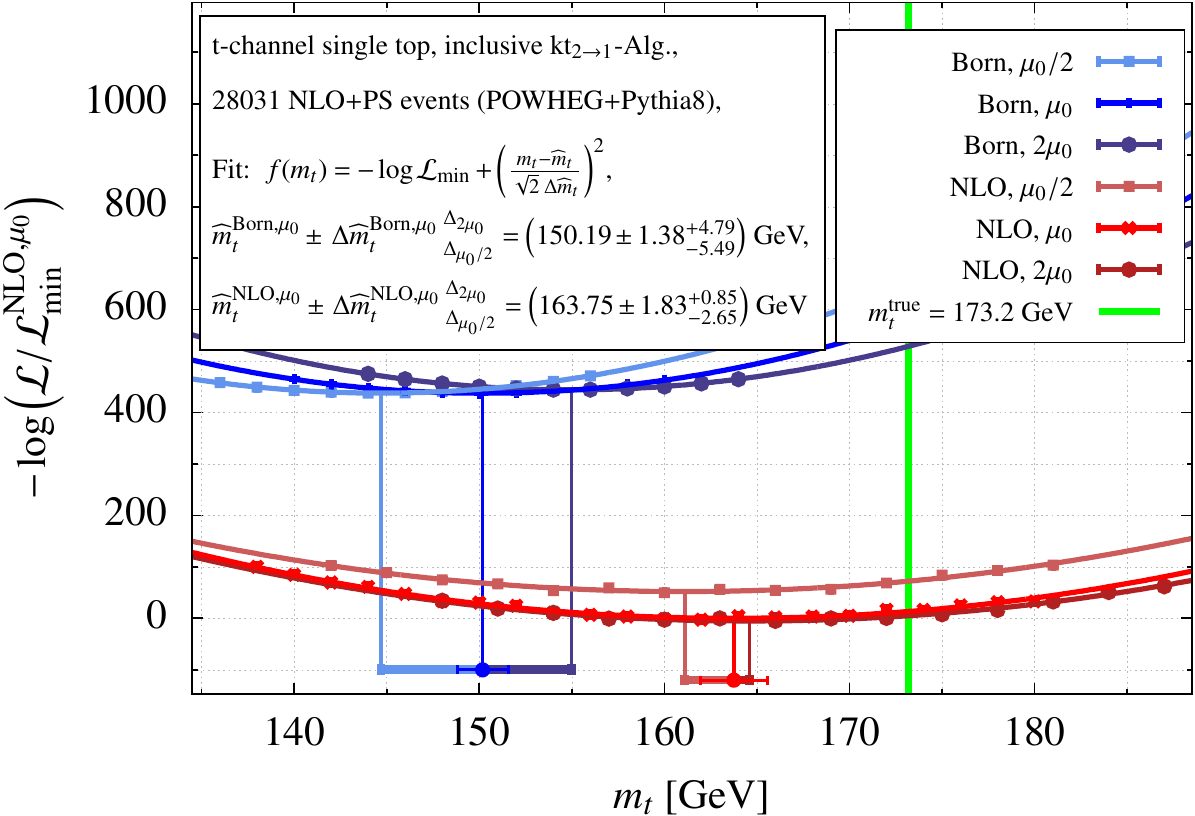}
  \includegraphics[width=0.497\textwidth]{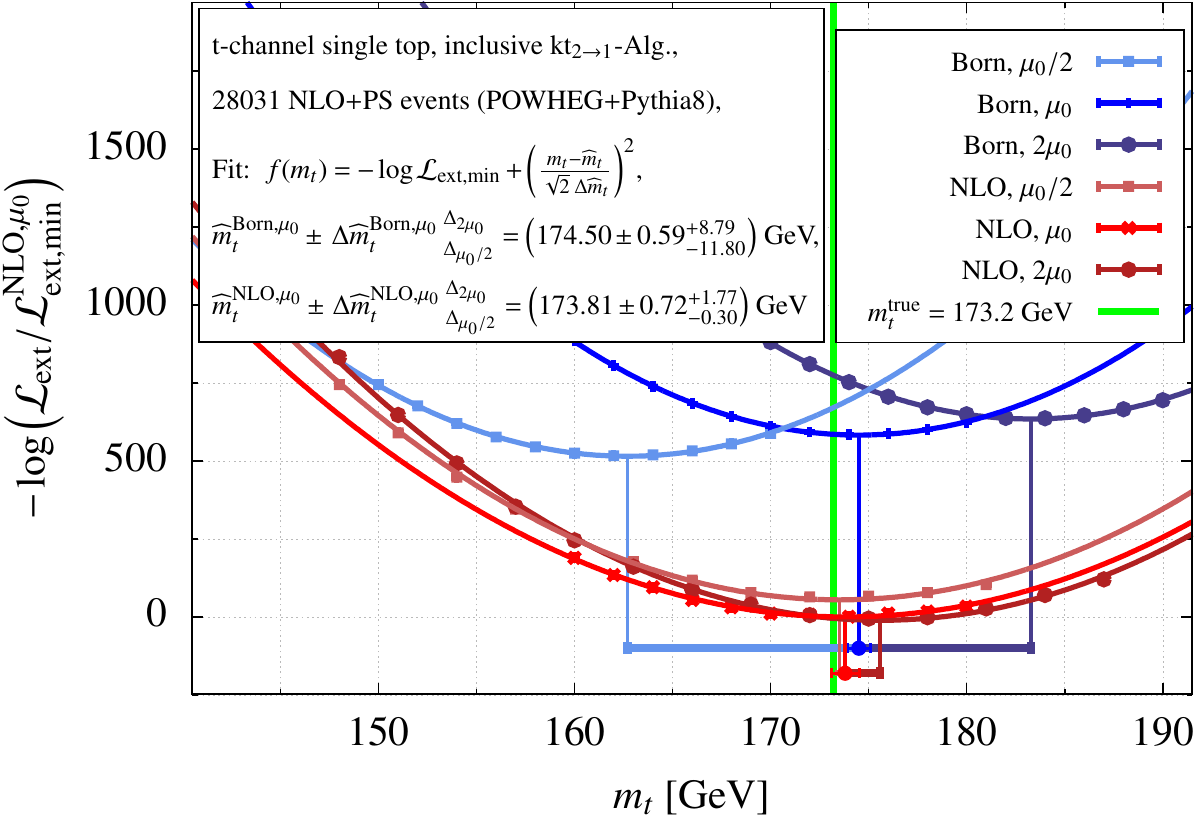}
\end{center}
\caption{{\bf Left: }Top-quark mass extraction via the MEM at LO (blue) and NLO (red) from single top-quark events generated with \textsc{POWHEG}+\textsc{Pythia8}. {\bf Right: }Same but with extended likelihoods functions.
}
\label{fig:memshower}
\end{figure}
The left plot of Fig.~\ref{fig:memshower} shows the extraction of the top-quark mass with the MEM from showered events generated with \textsc{POWHEG} and \textsc{Pythia8}. In the analysis shown in Fig.~\ref{fig:memshower} the likelihood function is calculated in the Born approximation only (blue) and including NLO corrections (red).  A parabola is fitted to the negative logarithm of the likelihood function ({\it Log-Likelihood}) in order to obtain the estimator for the top-quark mass $\hat{m}_t$ as the position of the minimum and the corresponding statistical uncertainty $\Delta\hat{m}_t$ as the width of the parabola. The theoretical uncertainty due to missing higher orders is estimated by varying the factorisation and renormalisation scales in the likelihood calculation by a factor of $2$. The biases in the extracted estimators $\hat{m}^{\mathrm{Born}}_t=150.19$~GeV and $\hat{m}^{\mathrm{NLO}}_t=163.75$~GeV with respect to the input value for the top-quark mass $m^{\mathrm{true}}_t=173.2$~GeV are caused by the missing higher-order corrections in the respective likelihood calculations: It is known that the MEM is prone to yield biased estimators in case of a mismatch between the probability density underlying the likelihood calculation and the actual distribution of the events. Thus, a calibration of the method is needed introducing its own associated uncertainties. Although, this bias, hence the amount of calibration, is reduced when NLO corrections are taken into account, both results are incompatible with the input value within their uncertainty bands. This means that the uncertainties due to missing higher-order corrections in general and the parton-shower effects in particular are not reliably estimated by the scale variation. 

The right plot of Fig.~\ref{fig:memshower} shows the top-quark mass extraction from the same set of events but using the {\it Extended Likelihood} defined by
\begin{displaymath}
\mathcal{L}_{\mathrm{ext}}(m_t|\{\bm{x}_i\})={\sigma(m_t)^N\cdot L^N\over N!}e^{-\sigma(m_t)\cdot L}\mathcal{L}(m_t|\{\bm{x}_i\})
\end{displaymath}
in the MEM. The normal likelihood function is combined with the Poisson probability of having observed $N$ signal events given the theoretically predicted number of events $\nu(m_t)=\sigma(m_t)\cdot L$ in terms of the fiducial cross section $\sigma$ and the integrated luminosity of the collider $L$. The parton shower modifies the phase space density by multiple parton emissions but does only affect the number of expected events mildly through acceptance effects. Thus, incorporating the information of the total number of recorded events via the {\it Extended Likelihood} at NLO accuracy gives a significant improvement of the sensitivity and accuracy of the analysis because the cross section exhibits an approximate linear dependence on the top-quark mass parameter \cite{Kant:2014oha}: As can be seen from the right plot of Fig.~\ref{fig:memshower}, not only the estimators get shifted but the statistical uncertainties are reduced by more than a factor of $2$ with respect to the previous analysis. The systematic uncertainty is also reduced in the case of the NLO-accurate {\it Extended Likelihood} but is enlarged by more than a factor of $2$ when only the Born approximation is utilised. In the analysis in the right plot of Fig.~\ref{fig:memshower} both estimators are compatible with the input value within their uncertainties, dispelling the need for calibration of the MEM and justifying to estimate the theoretical uncertainty by scale variation. Confidence in the validity of the estimated uncertainties is crucial to judge the precision of the MEM. More details can be found in Ref.~\cite{Kraus:2019qoq}.

\section{BSM parameter determination with the MEM@NLO}\label{sect:bsmmem}
As a second example, we study the sensitivity of parameter extraction with the MEM at NLO accuracy in a Beyond-Standard-Model scenario. We consider single top-quark production in association with a Higgs boson with a non-trivial mixing of CP-even and CP-odd top-quark Yukawa couplings. The coupling of the top quark ($t$) to the Higgs boson ($H$) is described by the effective Lagrangian
\begin{displaymath}
\mathcal{L}_{t\bar{t}H}=-{m_t\over v}\;\bar{t}\left(\cos\alpha+i\gamma_5{2\over 3}\sin\alpha\right)\;t\;H
\end{displaymath}
with the top-quark mass $m_t$ and the vacuum expectation value of the Higgs field $v$ \cite{Artoisenet:2013puc,Demartin:2015uha}.
The angle $\alpha$ parameterises the relative CP-mixing phase of the top-Higgs interaction with respect to the coupling of the Higgs boson to the $W$ boson. 
\begin{figure}[h!]
\begin{center}
  \includegraphics[width=0.75\textwidth]{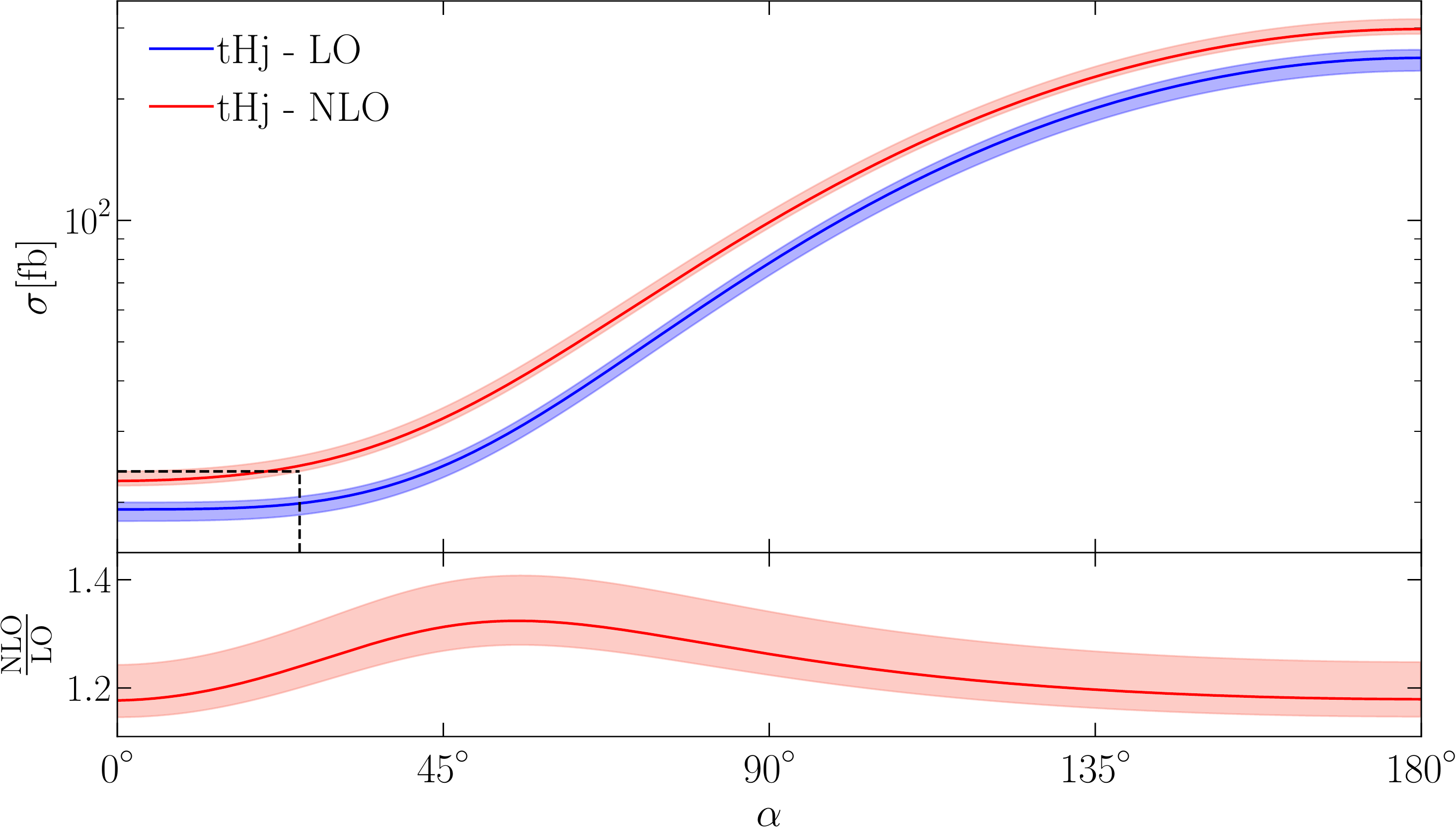}
\end{center}
\caption{The fiducial cross section of $pp \rightarrow tH$ as a function
of the CP-mixing angle $\alpha$ at LO and NLO accuracy.
}
\label{fig:sgthfidxs}
\end{figure}
In Fig.~\ref{fig:sgthfidxs} the cross section predicted at leading-order (blue) and NLO accuracy (red) is shown as a function of $\alpha$ together with theoretical uncertainty bands estimated by varying the renormalisation and factorisation scale by a factor of $2$. From the ratio NLO/LO, shown in the bottom panel, it is obvious that the NLO corrections are important and particularly dependent on $\alpha$. An interval in the parameter space of $a=0^\circ\leq\alpha\leq b\approx 25^\circ$ is implied by the dashed line. For $\alpha$ values in this interval the predictions for the cross section are compatible with the SM expectation within the theoretical uncertainties. An input value of ${\alpha}^{\tiny\textrm{in}}=22.5^\circ$ is chosen to generate events consisting of single top-quarks in association with a Higgs boson at NLO accuracy. 
\begin{figure}[h!]
\begin{center}
  \includegraphics[width=0.75\textwidth]{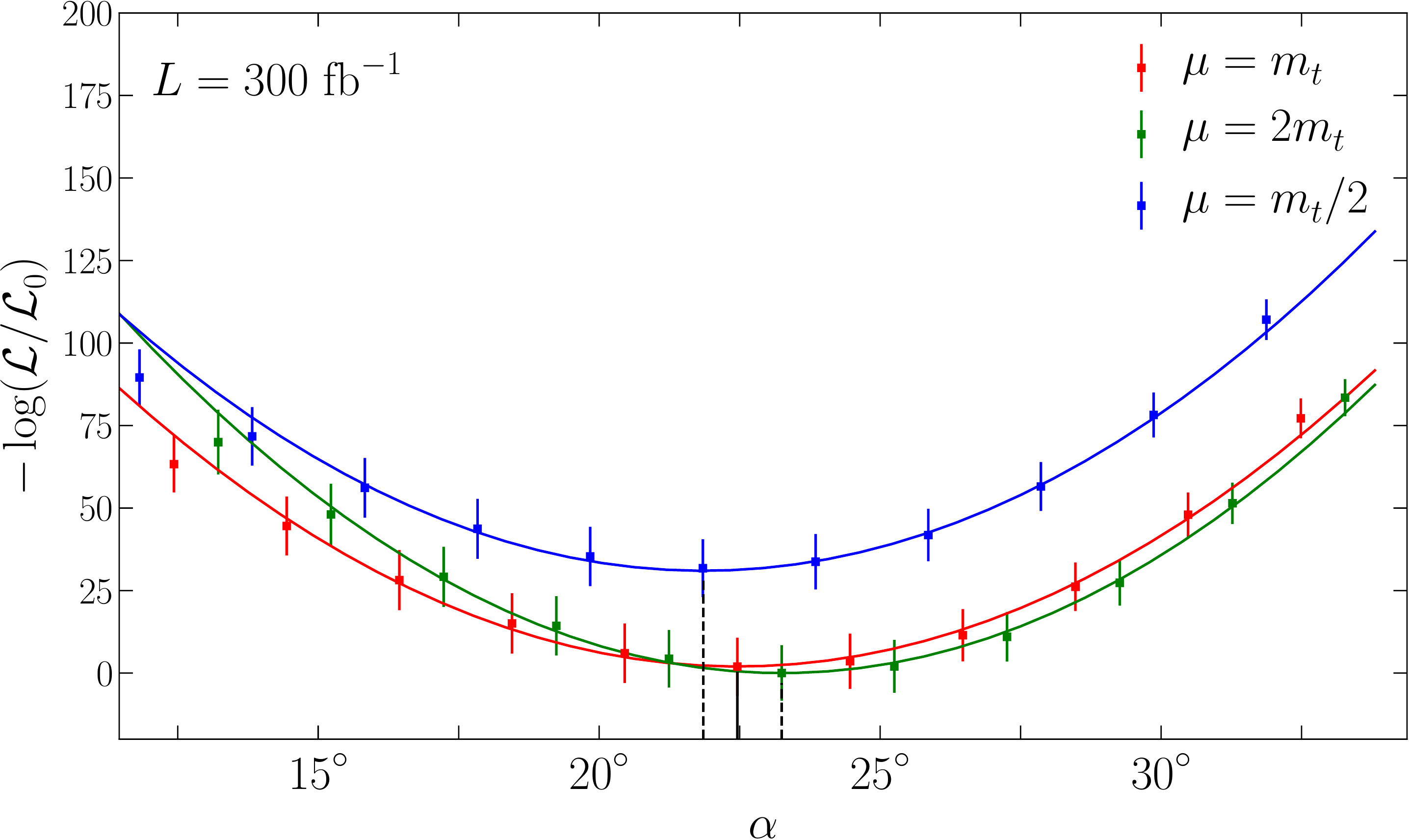}
\end{center}
\caption{Determination of the CP-mixing angle $\alpha$ with the MEM@NLO for $300\textrm{ fb}^{-1}$ of data.
}
\label{fig:memsgth}
\end{figure}
The MEM@NLO is employed to study the sensitivity of the BSM-parameter determination (see Fig.~\ref{fig:memsgth}). As expected, when using the same probability distribution in the MEM as in the event generation, the extracted estimator perfectly reproduces the input value. The statistical uncertainty can be derived from the width of the fitted parabola while the theoretical uncertainty is estimated by scale variation in the likelihood calculation. For an integrated luminosity of $L=300\textrm{ fb}^{-1}$ the analysis yields
$
\hat{\alpha}=22.5^\circ\pm0.9^\circ[\textrm{stat.}]^{+0.7^\circ}_{-0.6^\circ}[\textrm{sys.}]
$
with comparable statistical and systematic uncertainties. Note that no signal detection efficiencies and branching fractions are considered in this study. For more details the reader is referred to Ref.~\cite{Kraus:2019myc}.

\section{Conclusion}\label{sect:concl}
We have presented the MEM as a promising tool for precision measurements at the LHC once NLO corrections are consistently taken into account. The extension of the MEM to NLO accuracy is applicable to data defined by common jet algorithms. For the first time, it allows to study the impact of parton shower effects within the MEM. We have shown that incorporating NLO corrections in the calculation of the likelihood significantly reduces the bias stemming from missing higher-order effects by analysing showered events generated by \textsc{POWHEG}+\textsc{Pythia8}. In this way less calibration of the MEM is required but parton shower effects can still have a significant impact on the analysis. When considering the {\it Extended Likelihood} formalism the MEM@NLO yields unbiased estimators even when showered events are analysed with fixed-order predictions. In a second example we have shown that utilising the MEM@NLO to determine a BSM parameter can be beneficial compared to cross section measurements. We want to stress that in order to develop the MEM@NLO to be applicable to experimental data, the inclusion of more realistic final states (including e.g. decays), non-trivial transfer functions and parton shower effects in the theoretical predictions has still to be investigated in future studies.
\providecommand{\href}[2]{#2}\begingroup\raggedright\endgroup

\end{document}